\documentclass{emulateapj}

\usepackage{ graphicx}
\newcommand{\kms}{km~s$^{-1}$}

\begin{document}
\title{Cosmological concordance or chemical coincidence? Deuterated molecular hydrogen abundances at high redshift} 
\author{J. Tumlinson\altaffilmark{1}, 
A. L. Malec\altaffilmark{2}, 
R. F. Carswell\altaffilmark{3},
M. T. Murphy\altaffilmark{2}, 
R. Buning\altaffilmark{4}, 
N. Milutinovic\altaffilmark{5}, 
S. L. Ellison\altaffilmark{5}, 
J. X. Prochaska\altaffilmark{6}, 
R. A. Jorgenson\altaffilmark{3},
W. Ubachs\altaffilmark{4},
A. M. Wolfe\altaffilmark{7}}
\altaffiltext{1}{Space Telescope Science Institute, 3700 San Martin Drive, Baltimore, MD, 21218, USA}
\altaffiltext{2}{Centre for Astrophysics and Supercomputing, Swinburne University of Technology, Melbourne, Victoria 3122, Australia}
\altaffiltext{3}{Institute of Astronomy, University of Cambridge, Madingley Road, Cambridge CB3 0HA, UK}
\altaffiltext{4}{Laser Center, VU University, De Boelaan 1081 HV Amsterdam, The Netherlands} 
\altaffiltext{5}{Department of Physics \& Astronomy, University of Victoria, Victoria, BC V8P 1A1, Canada}
\altaffiltext{6}{University of California Observatories-Lick Observatory, UC Santa Cruz, CA 95064, USA}
\altaffiltext{7}{Department of Physics and Center for Astrophysics and Space Sciences, University of California, San Diego, 9500 Gilman Dr., La Jolla, CA 92093-0424, USA }
\begin{abstract}
We report two detections of deuterated molecular hydrogen (HD) in QSO absorption-line systems at $z > 2$. Toward J2123-0500, we find $N$(HD) $= 13.84 \pm 0.2$ for a sub-DLA with metallicity $\simeq 0.5Z_{\odot}$ and $N$(H$_2$) = $17.64 \pm 0.15$ at $z = 2.0594$. Toward FJ0812+32, we find $N$(HD) $= 15.38 \pm 0.3$ for a solar-metallicity DLA with $N$(H$_2$) = $19.88 \pm 0.2$ at $z = 2.6265$. These systems have ratios of HD to H$_2$ above that observed in dense clouds within the Milky Way disk and apparently consistent with a simple conversion from the cosmological ratio of D/H. These ratios are not readily explained by any available model of HD chemistry and there are no obvious trends with metallicity or molecular content. Taken together, these two systems and the two published $z > 2$ HD-bearing DLAs indicate that HD is either less effectively dissociated or more efficiently produced in high-redshift interstellar gas, even at low molecular fraction and/or solar metallicity. It is puzzling that such diverse systems should show such consistent HD/H$_2$ ratios. Without clear knowledge of all the aspects of HD chemistry that may help determine the ratio HD/H$_2$, we conclude that these systems are potentially more revealing of gas chemistry than of D/H itself and that it is premature to use such systems to constrain D/H at high-redshift.  
 \end{abstract}
\keywords{ISM: molecules --- quasars: absorption lines } 

\section{Introduction}

Absorption-line systems in the spectra of high-redshift sources such as QSOs and GRB afterglows provide opportunities to examine the interstellar gas in galaxies at early cosmic times with samples that are less biased than emission-selected samples. These techniques have been used to probe the properties of gas clouds at high redshift, where conditions can differ significantly from those in the local Universe. Detections of cold molecular gas at high redshift are less common than might be expected; only about one quarter of DLAs show detectable molecular hydrogen (Noterdaeme {et~al.} 2008a). Even in GRB-DLAs, where the sightline pierces deep inside the ISM of the host galaxy, molecular gas does not appear often (Tumlinson {et~al.} 2007), though it has been detected in one system via H$_2$ and CO absorption (Prochaska {et~al.} 2009). The opportunity to detect deuterated hydrogen molecules (HD) exists in many of these systems where H$_2$ is seen, and would in principle provide another useful constraint on physical conditions. HD is fragile, easily photo-dissociated, and is present in the Milky Way interstellar medium (ISM) only along reddened sightlines where the absorbing cloud is dense, dusty, and self-shielded. HD may also be an important gas coolant at low density in low-metallicity galaxies, and so it is important to understand the conditions in which it becomes abundant. In this Letter we report two new detections of HD in high-redshift QSO absorption line systems: the sub-DLA at $z = 2.09534$ toward J2123-0500 and the DLA at $z = 2.626$ toward FJ0812+32. These systems show H$_2$ and HD absorption but are otherwise quite different and revealing. 

\section{Data on J2123-0500}
\label{J2123section}

Data on J2123--0500 were obtained with Keck/HIRES in a series of exposures spanning the nights of 17-20 Aug 2006. A complete analysis of the spectrum is presented by Milutinovic et al. (submitted). The effective spectral resolution is $R \sim 100,000$ in the reduced spectrum, with S/N $\sim 15 - 20$ pix$^{-1}$ across the band where we search for molecular absorption. The sub-DLA at $z = 2.0594$ exhibits $\log N$(\ion{H}{1}) $ = 19.18 \pm 0.15$. These data have also been used to constrain the temporal evolution of the proton-to-electron mass ratio by Malec {et~al.} (2010, hereafter M10).  

\begin{figure}[ht]
\begin{center}
\includegraphics[width=3.3in]{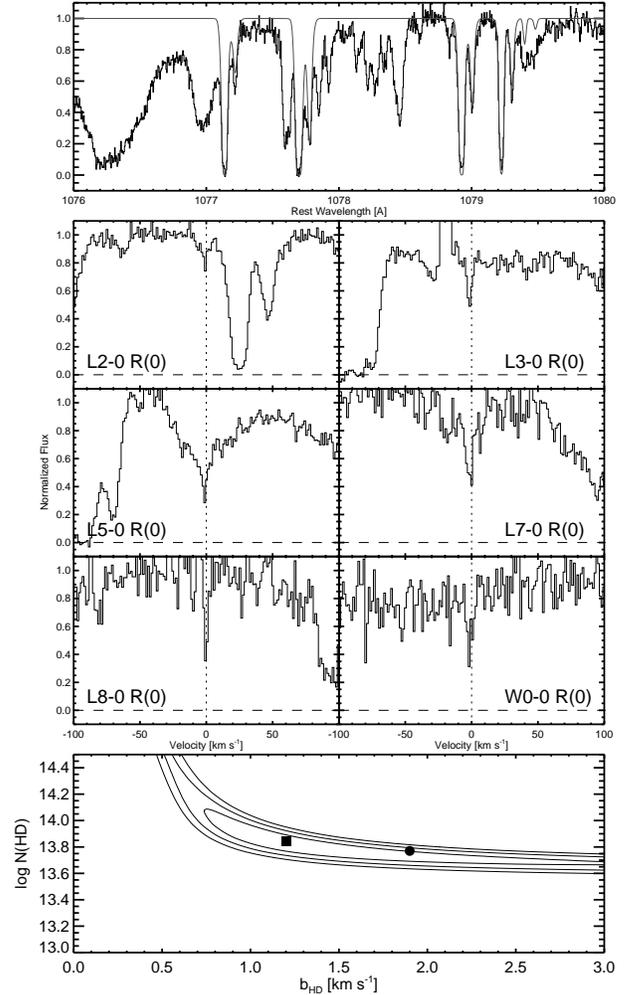}
\caption{H$_2$ and HD lines in the system at $z= 2.059345$ in the J2123-0050 sightline. The top panel shows $H_2$ lines from the Lyman 2-0 band with the best profile fit overlaid. The middle panels show six R(0) transitions of the HD Lyman-Werner bands. The lower panel shows the HD parameter space, with the best fits from the COG (filled square) and the M10 profile fit (filled circle). \label{hdlinefig1}} 
\end{center}
\end{figure}

We identified the HD lines shown in Figure \ref{hdlinefig1} using the laboratory wavelengths of Ivanov {et~al.} (2008) and the transition strengths of Abgrall \& Roueff (2006). Equivalent widths $W_{\lambda}$ (Table 1) were measured by integration over the line profile with respect to an iteratively fitted polynomial continuum. The detected lines were then fitted to a single-component curve of growth (COG) parameterized by HD column density $N$(HD) and the effective doppler width $b_{HD}$; the best fit was determined to be $\log N$(HD) $= 13.84$ and $b = 1.2$ \kms. Because the detected HD lines lie on the linear and knee portions of the COG, $b_{HD}$ is poorly constrained at the high end; conversely, the column density is well constrained to N(HD) $\geq 13.65$. As part of their global line-profile fitting, M10 obtained $\log N$(HD) $=13.77$ and $b = 1.9$ \kms, a consistent result (Figure~\ref{hdlinefig1}).

\begin{deluxetable}{lcccc}[t]
\tablecolumns{5} 
\tablenum{1} 
\tablewidth{0pt} 
\tablecaption{Summary of two HD systems} 
\tablehead{ \colhead{Line} 
&\colhead{$\lambda _0$ [\AA]\tablenotemark{1}}
&\colhead{$f$\tablenotemark{2}} 
&\colhead{$W_{\lambda}$ [m\AA]\tablenotemark{3}} & \colhead{$W_{\lambda}$ [m\AA]\tablenotemark{3}} \\
\colhead{} &  & \colhead{} & \colhead{J2123-0050} & \colhead{FJ0812+32B} 
} 
\startdata 
L0-0R(0)  & 1105.84 &  0.000743 & \nodata           & 7.8 $\pm$ 2.0\\
L1-0R(0)  & 1092.00 & 0.00297    & \nodata           & \nodata  \\
L2-0R(0)  & 1078.83 &  0.00675  & 4.4 $\pm$ 2.8 & \nodata \\
L3-0R(0)   & 1066.28 &  0.0115   & 5.6 $\pm$ 0.6 & 13.1 $\pm$ 2.2 \\
L4-0R(0)   & 1054.29 &  0.0164   & \nodata            & 13.4 $\pm$ 1.0\\
L5-0R(0)   & 1042.85 &  0.0206   & 8.0 $\pm$ 0.6 & 12.7 $\pm$ 2.2\\ 
L6-0R(0)   & 1031.91 &  0.0236   & \nodata             &  \nodata \\
L7-0R(0)   & 1021.46 &  0.0254   & 8.0 $\pm$ 1.2 & \nodata \\ 
L8-0R(0)   & 1011.46 &  0.0262   & 8.4 $\pm$ 1.3 & 13.8 $\pm$ 2.5\\
W0-0R(0) &  1007.29 &  0.0325   & 9.7 $\pm$ 2.6 & 18.4 $\pm$ 2.9\\
\cutinhead{Summary}
$\log N$(HI)   &  \nodata &  \nodata &   19.18$\pm$0.15 &     21.35$\pm$0.10                          \\
$\log N$(H$_2$) & \nodata & \nodata & 17.64$\pm$0.15 & 19.88$\pm$0.10\\ 
$\log N$(HD)        & \nodata & \nodata & 13.84$\pm$0.20 & 15.38$\pm$0.20  \\
$b_{HD}$ [km s$^{-1}$]      &  \nodata              & \nodata              &  1.9$^{+1.3}_{-0.5}$          &  0.85$\pm 0.20$
\enddata 
\label{line-list}
\tablenotetext{1}{Ivanov {et~al.} (2008)}
\tablenotetext{2}{Abgrall \& Roueff (2006)}
\tablenotetext{3}{Rest-frame}
\end{deluxetable} 

By contrast with the HD, there is considerable uncertainty in the H$_2$ column density. M10 assigned two velocity components to the HD-bearing absorber (1 and 2 in their Table 3); one of which is narrow and saturated in H$_2$. Their evidence for two components was based on careful treatments of the profile fits; ultimately two overlapping components are preferred by the data once continuum and zeropoint adjustments were taken into account (see their Section 3 and Figure 5). The column densities that minimize $\chi ^2$ for this component are N(0) = 15.80, N(1) = 17.52, for a total $\log N$(H$_2$) = 17.55 and $b_{H2} = 1.9$ km s$^{-1}$. These values yield an unphysical rotational temperature $T_{01} = 171 {\rm K} / \ln (9 N(0) / N(1) ) = -96$ K. Since these lines are highly saturated, much larger N(0) is possible with minimal change to the quality of the fit. Milutinovic et al. (2010) used only a single component for these profiles in their analysis of the metal lines in this system. 

\begin{figure}[ht]
\begin{center}
\includegraphics[width=3.5in]{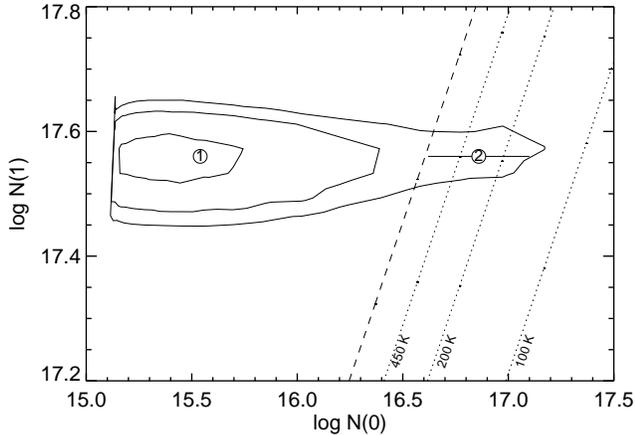} 
\caption{Contours of $\chi ^2$ for the saturated component of H$_2$ toward J2123-0500. Contours mark the 68, 95, and 99\% confidence intervals on total N(0) and N(1). Diagonal dotted lines mark the locations of $T_{01}$ = 100, 200, and 450 K, from right to left.  The dashed line marks the region where $T_{01}$ becomes undefined. Values left of this line are excluded on physical grounds. The two adopted column densities are marked with circled numerals.   
\label{h2contourfig}} 
\end{center}
\end{figure}

To explore how much higher N(0) could be in this saturated component, we searched a column-density space parameterized by N(0) and N(1). These results appear in Figure~\ref{h2contourfig}, where we plot contours of $\chi ^2$ for the global fit. For this test all other parameters in the global fit were allowed to settle at their optimal values. The point marked ``1'' is the fiducial fit adopted by M10, with N(0) = 15.56 and N(1) = 17.55. We find that the $\chi ^2$ contours extend up to N(0) $\simeq 17$ at 99\% confidence, well into the range where $T_{01}$ reaches reasonable physical values of 100 - 200 K. We therefore adopt a fiducial model that has N(1) = 17.55 from the M10 profile fit, and N(0) = $16.86 \pm 0.24$, as shown with the range bar in Figure~\ref{h2contourfig}. This latter value splits the difference between the 99\% confidence interval on N(0) and the line of unphysical $T_{01}$, and assigns an error that ranges to these extremes (point ``2''). This point is within the 99\% confidence interval for the profile fit and also gives a physically sensible $T_{01} = 281$ K (with a lower bound of 147 K). This temperature corresponds to a thermal $b_{H2} = 1.5$ km s$^{-1}$ (consistent with the profile-fitted $b = 1.9$ \kms) and $b_{HD} = 1.3$ km s$^{-1}$ (consistent with the COG-derived value of 1.2). This model is consistent with the available line strengths and widths for both H$_2$ and HD and becomes our fiducial model. The total $\log N$(H$_2$) $= 17.64 \pm 0.15$ and we find $\log N$(HD) / 2N(H$_2$) $ = -4.1 \pm 0.2$. 

We attempted a model of this absorber with a single velocity component in the H$_2$ using a COG; for this case N(0) =  15.70, N(1) = 16.20, and $\log N$(H$_2$) = 16.41. This fit has the advantage of not requiring any assumption about the $J=0$ column density, but has the disadvantage of being an inferior fit to the data, which strongly prefer a two-component fit (M10). This COG has doppler $b_{H2} = 4.6$ km s$^{-1}$, which does not agree as well with the $b_{HD} = 1.9$ km s$^{-1}$ as the profile fit above. Most significantly, this much lower N(H$_2$) yields a highly anomalous ratio of HD / 2H$_2$ $\sim 10^{-3}$. We report this ``low H$_2$'' model here and include this value in later analysis for completeness. 

\begin{figure}[ht]
\begin{center}
\includegraphics[width=3.3in]{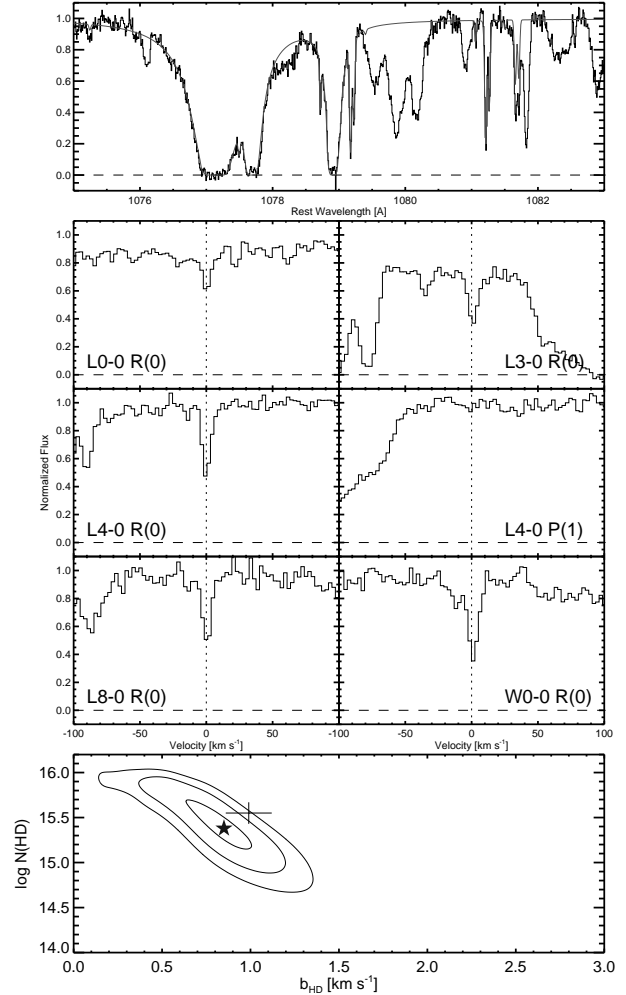} 
\caption{H$_2$ and HD lines in the DLA at $z= 2.626$ toward FJ0812+32. The top panel shows H$_2$ lines from the Lyman 2-0 band with the best profile fit overlaid. The middle panels show five R(0) transitions and one P(1) transition in the HD Lyman-Werner bands. The lower panel shows the HD parameter space and the best fit from the COG (filled star). The profile-fitted values are marked with the cross. \label{hdlinefig2}} 
\end{center}
\end{figure}

We do not detect $J=1$ HD in this system, but we can set a limit. The 5-0R(1) line at 1043.295 \AA\ falls into a region of clean continuum with good S/N. We obtain $W_{\lambda} \leq 2$ m\AA\ ($3\sigma$), which translates to N(1) $\leq 1.5 \times 10^{13}$ cm$^{-2}$. From this and the measured N(0), we infer $T_{01} \leq 46$ K for this absorber. This temperature is lower than the (problematic) $T_{01} = $ 292 K for the H$_2$. In interstellar H$_2$, $T_{01}$ is regarded as a reliable measure of the gas temperature only above a density $n_H \sim 100$ cm$^{-3}$ at which collisions effectively populate $J=1$ from $J=0$ via proton interchange reactions (Dalgarno, Black, \& Weisheit 1973). Since HD has a dipole moment, radiative transitions between $J=1$ and $J=0$ are permitted (as they are not for symmetric H$_2$) and HD can take on a lower excitation temperature than H$_2$ even at high density. Thus this measurement is probably not a reliable indicator of temperature in the HD absorbing cloud.  

Of the Lyman series, only Ly$\alpha$ and Ly$\beta$ are covered by our data at $\geq 989$ \AA\ in the rest frame for $z = 2.09534$. The Ly$\alpha$ and Ly$\beta$ profiles are damped, and the expected \ion{D}{1} profile at $-82$ \kms\ cannot be measured. 

\section{Data on FJ0812+32B}
\label{FJ0812section}

The data on FJ0812+32 were obtained with Keck/HIRES in a series of exposures totaling 14,000 s over 16-17 Mar 2005. Another exposure was obtained on 12 Jan 2008 with the same setup but bluer wavelength coverage. The data were reduced using standard XIDL reduction packages. The effective spectral resolution is $R \sim 50,000$ in the reduced spectrum, with S/N $\sim 40$ pix$^{-1}$ across the band where we search for H$_2$ and HD, declining to $\sim 20$ pix$^{-1}$ in the blue. These high-quality data have previously revealed a metal-strong DLA (Prochaska, Howk, \& Wolfe 2003) with an extremely cold gas cloud  Jorgenson {et~al.} (2009, hereafter J09).  

This system shows six detected HD R(0) lines; the other lines missing from Table 1 are blended with other features. These lines 
fall onto the flat portion of a single-component COG (Figure~\ref{hdlinefig2}), where their column density is only moderately constrained. However, the effective dopper $b_{HD}$ parameter is well-constrained; the best fit is $\log N$(HD) $= 15.38 \pm 0.25$ and $b_{HD} = 0.85 \pm 0.20$ \kms\ from the COG. A profile fit using VPFIT\footnote{http://www.ast.cam.ac.uk/$\sim$rfc/vpfit.html} obtained $\log N$(HD) $= 15.50$ and $b_{HD} = 1.0$, as shown in Figure 3. The column densities of H$_2$ in this system were also obtained as part of this global profile fit, yielding N($J$) = 19.81, 19.13, 16.20, 14.86, and 13.97 for $J =0$, 1, 2, 3, 4, respectively, for $\log$ N(H$_2$) $= 19.88 \pm 0.1$. The best-fit doppler  $b$ was held at $1.0$ \kms\ for $J = 0$ and 1 under the assumption that the bulk of the column density in H$_2$ coincides with the cold component of carbon and so is thermally broadened at $115$ K. These linewidths are consistent with the COG and profile-fitted values for HD at this temperature. We therefore conclude that the HD in this cloud plausibly arises in the same cold gas component as the cold C and H$_2$ (J09). 

There is also tentative evidence for HD $J=1$ in this system. Three $J=1$ lines appear as excess absorption in blended profiles of unrelated absorption; these are Werner 0-0R(1) and Lyman 5-0R(1) and 4-0P(1). Only the Lyman 4-0P(1) line at 1055.56 \AA\ is isolated and is marginally detected (see Figure 3). Including all four lines in the global fit yields N(1) $= 13.50 \pm 0.12$ for $b_{HD} = 1.0$ \kms\ fixed at the same value as for HD $J=0$. If we use this tentative value for N(1) we obtain $T_{01} = 26$ K, colder than the corresponding C I and H$_2$, for which J09 inferred $T < 74$ K. As for J2123-0500, we attribute this low excitation temperature to efficient radiative decay between $J=1$ and $J=0$ rather than the gas kinetic temperature. 

\section{Comparisons and Interpretations}
\label{interpsection}

We are now ready to compare these detections to other known HD-bearing gas. HD is known to be a component of some dense Galactic clouds from detections in sightlines to bright stars in the disk. This population is represented in Figure 4 by the sample of Lacour {et~al.} (2005), who reported HD toward 17 Galactic OB stars (10 from Copernicus, 7 from FUSE). These stars have E(B-V) = $0.1 - 0.7$, with molecular fractions $f_{H2} = 0.06 - 0.7$. 

\begin{figure}[!t]
\begin{center}
\includegraphics[width=3.5in]{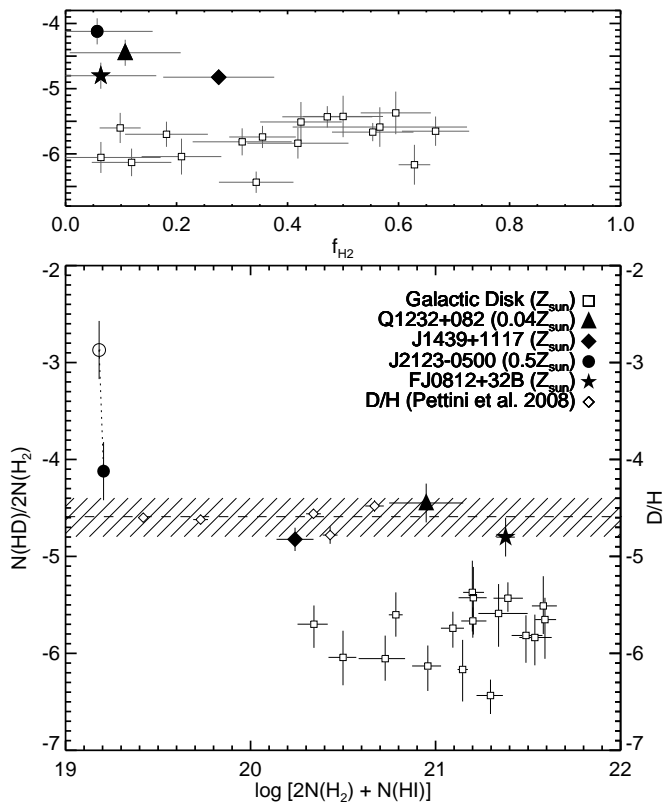} 
\caption{Comparison of the our two new HD abundances (filled circle for J2123-0500 and filled star for FJ0812+32B) relative to H$_2$ in the Galactic disk from Lacour {et~al.} (2005) (open squares), and two DLA systems 
from Varshalovich {et~al.} (2001, filled triangle), and Noterdaeme {et~al.} (2008b, filled diamond).  The cross-hatched area refers to the axis at right and marks the CMB-derived D/H ratio from Dunkley {et~al.} (2009). \label{hdplot}} 
\end{center}
\end{figure}

HD has been reported previously in two DLA systems. Varshalovich {et~al.} (2001) and, later, Ivanchik {et~al.} (2010) reported a detection of HD with N(HD) $= 3.4^{+1.6}_{-0.8} \times 10^{15}$ cm$^{-2}$ and N(H$_2$) = $4.8 \pm 1.0 \times 10^{19}$ cm$^{-2}$ in the DLA at $z =  2.3377$ toward Q1232+082, which has $N$(HI) = 20.90 $\pm$ 0.08. Also, Noterdaeme {et~al.} (2008b) have detected HD in a $z = 2.418$ DLA toward SDSS J1439+1117 with  $\log$ N(H$_2$) $= 19.38 \pm 0.10$,  $\log N$(HD) = $14.87 \pm 0.025$, and $N$(HI) = 20.1 $\pm$ 0.1. This HD to H$_2$ ratio is also well above the Galactic disk ratio. 

We have plotted two symbols in Figure~\ref{hdplot} for the DLA toward J2123-0500: the open symbol uses the lower COG-fitted measurement for N(H$_2$), while the filled circle uses the higher profile fitted value (\S~2). The error bars on the points express the estimated {\it statistical} noise on each measurement; the line connecting them shows the larger systematic uncertainty arising from saturation of the H$_2$ lines. With the higher H$_2$ column density, the HD to H$_2$ ratio in this system more closely resembles the other systems and the cosmological ratio of D/H. 

Our findings in the J2123-0500 and FJ0812+32B systems agree in their HD/H$_2$ ratio with the other detections, with the striking result that all four high-redshift systems lie both well above the Galactic disk value and apparently consistent with the cosmological D/H ratio. This finding is somewhat puzzling, in light of the diversity in the other properties of these systems. In their analysis of Q1232+082, Ivanchik {et~al.} (2010) explain this effect by low astration at early times. However, these systems show metallicities proxied from S or Zn that are between 0.04 $Z_{\odot}$ and $Z_{\odot}$ (as shown in the legend of Figure~\ref{hdplot}). It is puzzling why the infall of gas undepleted in D would be arranged just so that the primordial D/H ratio is maintained in four such diverse systems. It is not clear why low astration would be the dominant factor in environments that have already achieved a metallicity comparable to the Milky Way disk, as it is for three of the systems. Indeed, the consistency of the HD/H2 ratio across the factor of 20 in metallicity spanned by these systems indicates that metallicity is perhaps only a minor factor in determining the abundance of HD. 

Even with four systems that lie along a narrow band of HD/H$_2$ ratio well above the Galactic ratio, we cannot be sure that this is the general pattern of HD/H$_2$ in high-redshift galaxies. HD may have gone undetected in other DLAs where the Galactic ratio would place the HD below detection limits. Our column density sensitivity limits here are N(HD)  $\simeq 10^{13}$ cm$^{-2}$, which gives a limit HD/2H$_2$ $= 1.1\times10^{-5}$ for J2123-0500 and $6.6 \times 10^{-8}$ for FJ0812+32B, well below the Galactic points in the latter case but above them in the former. Though in most published cases the data quality is adequate to detect HD in the ratios seen here, a systematic survey of H$_2$-bearing DLAs might reveal that in some cases a Galactic ratio is permitted given the individual detection limits. In light of this ``detection bias'' we cannot yet conclude that the anomalous ratios of the four systems known so far are representative of the general pattern at high redshift. 

Another striking feature of these systems is that their typical molecular fraction is $f_{H2} \sim 0.06 - 0.3$. The typical value for the Galactic disk clouds is up to ten times higher, with most scattering in between $f_{H2}$ = 0.2 and 0.6 (Figure~\ref{hdplot}). In the Galactic disk, the HD is believed to occupy only the dense, self-shielded inner regions of the cloud and to be dissociated easily in the low-extinction photodissocation regions surrounding the cloud (Lacour {et~al.} 2005). By this line of reasoning, clouds with lower molecular fractions should have more of their HD dissociated and should lie below the Galactic disk points rather than above. This effect is particularly notable for the J2123-0500 system, which does not qualify as a DLA and has the lowest $f_{H2} = 0.028$. The mystery is then deepened, in the sense that we must explain a higher HD to H$_2$ ratio than in the Milky Way in clouds of similar metallicity but lower total molecular abundance. 

Dust is another factor that may help promote the formation of HD in these systems.  All four of these HD-bearing systems show evidence of dust depletion based on dust-sensitive abundance ratios. The system toward J2123-0500 shows ionization-corrected abundance ratios Si and Fe depleted by approximately 0.5 dex with respect to S and N (see table 4 of Milutinovic et al.). The F0812+32B system shows solar [S/Si] but [Zn/Fe] and [Cr/Zn] that indicates a high depletion of Fe (J09). The 2175 \AA\ dust feature is not detected in either spectrum. Unfortunately, beyond these abunance ratios no more quantitative information has been obtained on the exact dust-to-gas ratios or dust properties of these two systems, so it is not known whether the composition and size distribution of the dust is similar to that of the Galactic disks. We therefore cannot correlate the HD properties of these systems against dust in any more detail. 

Using a master-equation approach, Lipshtat, Biham, \& Herbst (2004) studied the relative formation rates of H$_2$ and HD under the assumption that D atoms stick to dust grain surfaces for slightly longer than H. They have found that for a certain range of conditions, the total formation rate of HD can rise from $\sim 10^{-5}$ to $\gtrsim 10^{-3}$ times that of H$_2$ as the dust temperature rises from $< 15$ K to 20 K. The effect is strongest for small grains with a small count of binding sites to which H and D attach. 

Since this model describes the formation rates but not the relative photo-dissociation rates (which are greatly influenced by self-shielding for H$_2$ and much less so for HD), and since it possesses at least three free parameters for density, grain size, and grain temperature, we cannot directly use our data to constrain these models. We can speculate that small grains, the right range of dust temperatures, or both, may together establish the observed molecular abundance ratios seen in these four $z > 2$ systems. However, with the relative rates of H$_2$ and HD production ranging over a factor of 100 between $T_{dust} \leq 15$ K and $20$ K for a range of grain sizes, it is not clear why the four systems found so far have scatter so closely around the cosmological D/H. 

We also compare our HD results against the D/H ratios seen in high-redshift DLAs as compiled by Pettini {et~al.} (2008), represented by open diamonds in Figure~\ref{hdplot}. We note that the D/H measurements are quite precise and show a low scatter consistent with the CMB-derived value. In light of the precision and consistency of these measurements, it seems unnecessary to rely on HD/H$_2$ as a measure of cosmological D/H when the chemical influences on HD are so uncertain. 

In light of the diversity of these systems in metallicity and molecular fraction, and with the discussion above in mind, we conclude that the balance of chemical reactions, dust properties, and local conditions are highly uncertain in these systems. Following this, it seems that  these systems pose a puzzle about interstellar chemistry more effectively than they solve one about D/H. If properly understood these systems are likely to be far more revealing of these uncertain chemical and environmental effects than they are of the primordial abundances, which in any case appear well constrained from direct measurements of D/H. We can hope the discovery of more such systems and additional theoretical work will be able to disentangle these effects and resolve these doubts. 

\acknowledgements
A portion of this work was completed while J. T. was supported by the Gilbert and Jaylee Mead Postdoctoral Fellowship in the Yale Department of Physics. M. T. M. thanks the Australian Research Council for a QEII Research Fellowship (DP0877998). J. X. P. is supported by NSF grants AST-0548180 and AST-0709235. R. F. C. is grateful to the Leverhulme Trust for an Emeritus award. We are grateful to an anonymous referee for constructive comments that improved the manuscript. 

% astronat stuff 
% run this first to process the output of Papers
%     $home/bibtexformat/bibtexformat hdpaper.bib -s -pn -f -auth 4 -abb -sort -o test.bib
%\bibliographystyle{/Users/jason/texmf/astronat/apj/apj}
%% \bibliography

\end{document}